Comment on
"Radiative effect on natural convection flows in porous media",
A.A. MOHAMMADEIN, M. A. MANSOUR, SAHAR M. ABD EL GAIED and RAMA SUBBA REDDY GORLA [Transport in Porous Media **32**:263-283, 1998]

Asterios Pantokratoras
Associate Professor of Fluid Mechanics
School of Engineering, Democritus University of Thrace,
67100 Xanthi – Greece
e-mail:apantokr@civil.duth.gr

In the above paper the authors treat the natural convection boundary layer flow in a Darcy-Brinkman-Forchheimer porous medium. In the energy equation the radiation effect has been taken into account. A two-parameter perturbation method is used for the solution of the equations. The first order results are presented in tables and figures. This is an interesting work but there are some weak points which are presented below:

Velocity and temperature profiles are presented in 16 figures and each figure caption has the form " Velocity or temperature profiles ...... for various values of R and $C_T$" where R is a dimensionless radiation parameter and $C_T$ is a temperature ratio. However all figures contain values of R but none of them contains any value of $C_T$. In addition the parameter $C_T$ is defined in nomenclature as temperature ratio but no further definition is given. What temperatures?

In pages 267 and 268 the boundary conditions for velocity and temperature profiles are given at η=0 and ∞ where η is the transverse similarity variable. The common feature in these boundary conditions is that velocity and temperature become zero as η → ∞. It is known in boundary layer theory that velocity and temperature profiles approach the ambient fluid conditions asymptotically as η → ∞ and do not intersect the line which represents the boundary conditions. In figure 1 we show a temperature profile taken from figure 8 of the above work. We see that the temperature profile does not approach the ambient temperature asymptotically but intersects the horizontal axis almost vertically. At the same figure we show the correct shape of this temperature profile. It is clear that this temperature profile given by A.A. Mohammadein et al. (1998) is wrong and this happens in many figures. One velocity profile included in figure 1, four velocity profiles in figure 2, one temperature profile in figure 3, six temperature profiles in figure 4, one velocity profile in figure 5, three velocity profiles in figure 6, one temperature profile in figure 7, three temperature profiles in figure 8, one velocity profile in figure 9, three velocity profiles in figure 10 and four temperature profiles in figure 12 intersect the horizontal axis and are wrong. Except that figure 3 is completely wrong because $\varphi_0(0)$ must be 1 and not 100. It is clear that the profiles which do not approach the horizontal axis asymptotically and intersect it, are truncated due to a small calculation domain used. The authors used for all the above cases a calculation domain with $\eta_{max}=5$. However this calculation domain was not sufficient to capture the real shape of profiles and a wider calculation domain, greater than 5, should be used as happens in figures 13,14, 15 and 16 where $\eta_{max}=12$. It is sure that the truncation of the above profiles has introduced errors and in the data included in tables I, II, III, and IV of the above paper.

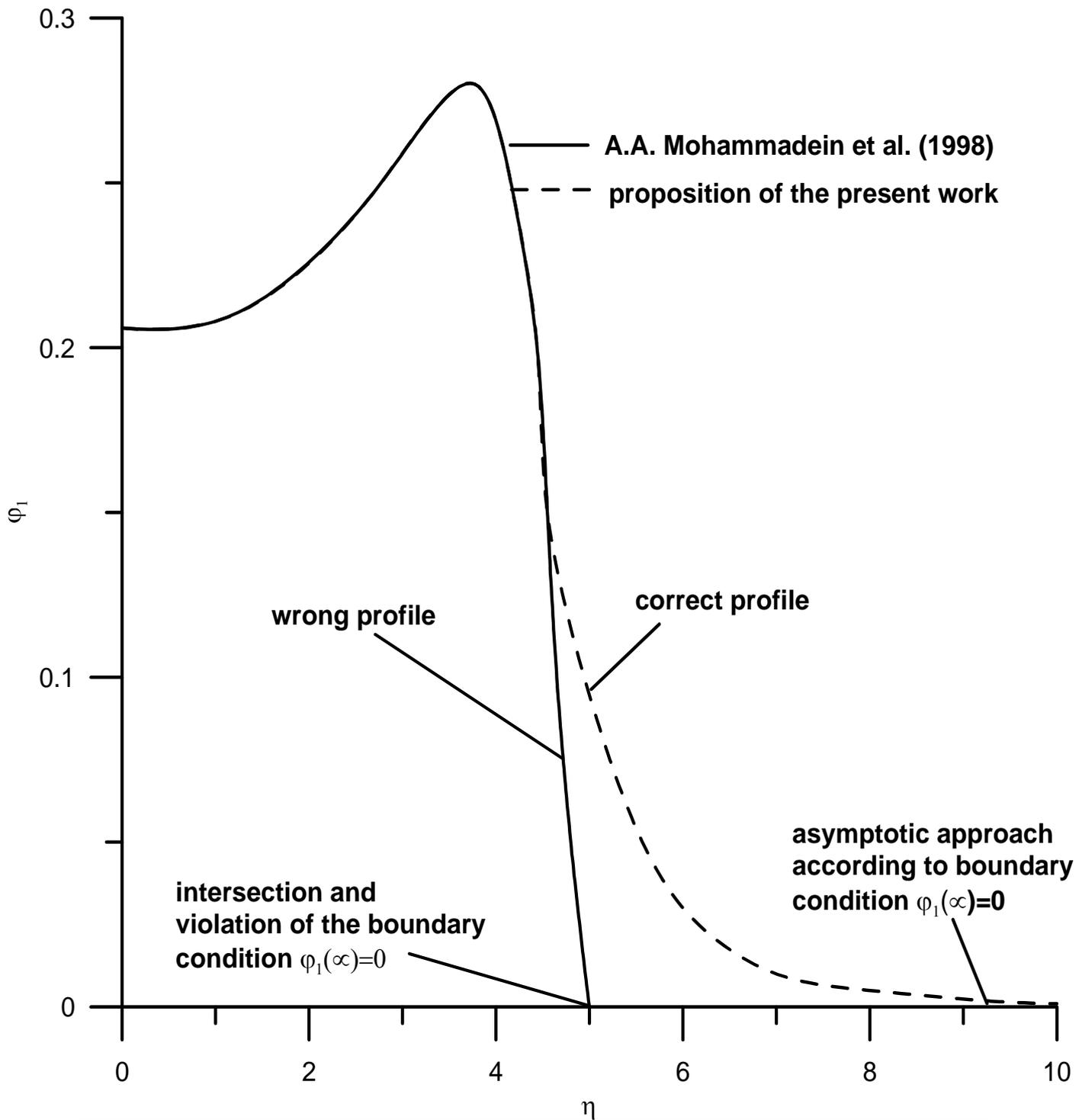

Figure 1. The solid line represents a dimensionless temperature profile for R=10. This profile has been reproduced from figure 8 by A.A. Mohammadein et al. (1998).

# REFERENCES

1. A.A. MOHAMMADEIN, M. A. MANSOUR, SAHAR M. ABD EL GAIED and RAMA SUBBA REDDY GORLA, Radiative effect on natural convection flows in porous media, Transport in Porous Media **32**:263-283, 1998.